\documentstyle [12pt]{article}

\begin{document}

\vspace{2mm}

\begin{flushright}
Preprint MRI-PHY/15/94, TCD-9-94 \\

hep-th/9411076, October 1994
\end{flushright}

\vspace{2ex}

\begin{center}

{\large \bf Singularities in Graviton-Dilaton System:

\vspace{2ex}

            Their Implications on the PPN Parameters

\vspace{2ex}

            and the Cosmological Constant } \\

\vspace{6mm}
{\large S. Kalyana Rama}
\vspace{3mm}

Mehta Research Institute, 10 Kasturba Gandhi Marg,

Allahabad 211 002, India.

\vspace{1ex}
email: krama@mri.ernet.in   \\
\end{center}

\vspace{4mm}

\begin{quote}
ABSTRACT.  Alternatives to Einstein's theory of general relativity
can be distinguished by measuring the parametrised post Newtonian
parameters. Two such parameters $\beta$ and $\gamma$, equal to
one in Einstein theory, can be obtained from static spherically
symmetric solutions. For the graviton-dilaton system, as in Brans-Dicke or
low energy string theory, we find that if $\gamma \ne 1$ for a charge
neutral point star,
then there exist naked singularities. Thus, if $\gamma$
is measured to be different from one, then it cannot be explained by these
theories, without implying naked singularities.
We also couple a cosmological
constant $\Lambda$ to the graviton-dilaton system, a la string theory.
We find that static spherically symmetric solutions in low energy string
theory, which describe the gravitational field of a point
star in the real universe atleast upto a distance
$r_* \simeq {\cal O} ({\rm pc})$, always lead to
curvature singularities. These singularities are stable and
much worse than the naked ones. Requiring their absence
upto a distance $r_*$ implies a bound
$| \Lambda | < 10^{- 102} (\frac{r_*}{{\rm pc}})^{- 2}$ in natural units.
If $r_* \simeq 1 {\rm Mpc}$ then
$| \Lambda | < 10^{- 114}$, and if $r_*$ extends all the way upto the edge
of the universe ($10^{28} {\rm cm}$) then
$| \Lambda | < 10^{- 122}$ in natural units.

\end{quote}

\newpage

\vspace{2ex}

\centerline{\bf 1. Introduction}

\vspace{2ex}

In Einstein's theory of general relativity, the gravitational field of a
point star is described by the static, spherically symmetric
Schwarzschild solution. Its predictions
have been verified to a very good accuracy. However, for various reasons,
as described in detail in \cite{will}, it is worthwhile to consider
alternative theories of gravity. Among the popular ones are the
Brans-Dicke (BD) theory, which is parametrised by a constant
$\omega > 0$, and the string theory. A common feature among these
generalised theories is the presence of a scalar field $\phi$, called BD
scalar or dilaton. There are other generalisations of BD theory, where
$\omega$ is a function of $\phi$, or the matter couplings to gravity
depend on another function of $\phi$, etc.\ . For details
see \cite{will}. We will consider here only BD theory and the low
energy limit of the string theory.

These alternative theories can be distinguished by measuring a set of
parameters called parametrised post Newtonian (PPN) parameters. Two such
parameters, $\beta$ and $\gamma$ can be obtained from static
spherically symmetric solutions of the graviton-dilaton system.
In Einstein's theory $\beta = \gamma = 1$.
Experimentally, their measured values are given by
$\frac{1}{3} (2 + 2 \gamma - \beta) = 1.003 \pm .005$ and
$\gamma = 1 \pm .001$. The
parameter $\beta$ is a measure of non linearity in the superposition law
for gravity, and $\gamma$ is a measure of the space time
curvature \cite{will}.
In this paper, we study the static spherically symmetric
solutions for BD and string theory,
including only the graviton and the dilaton field. They
describe the gravitational field of a point star, in these theories.
We find that the only acceptable solutions all lead to the
same predictions for the values of $\beta$ and
$\gamma$ as in Einstein's theory,
namely $\beta = \gamma = 1$. There are more general
static spherically symmetric solutions \cite{b}-\cite{tcd6}, which
predict $\beta = 1, \; \gamma = 1 + \epsilon$.
However, these solutions
always have naked curvature singularities proportional to
$\epsilon^2$ and,
hence, are unacceptable.

These general
solutions can be better understood by coupling the electromagnetic
field \cite{gm,ghs}. They lead to non trivial PPN parameters for
a point star of mass $M$ and charge $Q$ . In these solutions there is
an inner and an outer horizon. The curvature scalar is singular
at the inner horizon, but this singularity is
hidden behind the outer horizon. A charge neutral star can
then be obtained in two ways: in one, corresponding to the Schwarzschild
solution, the PPN parameters are trivial and
there is no naked singularity, while in the other, the PPN parameters
are non trivial but there is a naked singularity.

Therefore neither BD nor low energy string theory can predict non
trivial values for
PPN parameters $\beta$ and $\gamma$, for a charge neutral star,
without introducing naked singularities.
Thus if naked singularities are forbidden then, for a charge
neutral star, both the BD and the low energy string theory lead to the
same predictions for $\beta$ and $\gamma$ as in Einstein theory. In
particular, if the parameter $\gamma$ for a charge neutral point star is
known to be different from one, then it cannot be explained by
either BD or low energy string theory, without implying the existence of
a naked singularity. In that case an alternative
theory is needed that can predict a non trivial value for $\gamma$
for a charge neutral star, without any naked singularity.

In Einstein's theory, one can also add a cosmological constant $\Lambda$.
The only modifications to the static spherically symmetric solutions are
that, the space time is not flat asymptotically and, for $\Lambda > 0$, it
develops a new cosmological horizon \cite{gh}.
In the second half of this paper, we couple
the cosmological constant to the dilaton $\phi$, in a way
analogous to the coupling of a tree level cosmological constant in low
energy string theory \cite{tdgang}.

The time dependent, expanding universe type of solutions to such system
have been extensively studied in various space time dimensions
\cite{tdgang}. In the low energy limit of the string theory, the
dilaton is expected to develop a potential and acquire a mass. Hence,
$\Lambda$ can be considered as a function of $\phi$, leading to a dilaton
potential. Taking it to be of the form
$e^{- \phi} \Lambda (\phi) = m (\phi - \phi_0)^2 + \cdots$
near the minimum, the authors of \cite{hh} have thoroughly analysed
the implications of such a massive dilaton for the static spherically
symmetric case. Depending on the choice of
$m$, the system is expected to develop one, two, or three horizons.
Also the static solutions to $d + d_i + 2$ dimensional
gravity with a higher dimensional cosmological constant have been studied
in \cite{wilt}, where $d_i$ is the number of internal dimensions.

In this paper, we take $\Lambda$ to be a constant and analyse the static
spherically symmetric solutions in $d = 4$ space time.
They describe the gravitational
field of point stars, and continue to do so to a very good
approximation even when the stars have non relativistic velocities with
respect to each other. For example, the Schwarzschild solution describes
very well the gravitational effect of the sun on earth even though
the earth is revolving around the sun with a speed
of ${\cal O} (10 {\rm km/sec})$. Also, in Einstein's theory, when a
cosmological constant $\Lambda$ is present, the static spherically
symmetric solutions still describe the gravitational field of
point stars and also the redshift of distant objects \cite{wein}.
Therefore one expects that in our present case also,
when a cosmological constant $\Lambda$ is present, the
static spherically symmetric solutions will describe the gravitational
field of stars atleast upto a distance $r_*$, even though the real
universe is not static but expanding, characterised by the
Hubble constant $H_0 = 100 h_0 {\rm km/sec/Mpc}, \; 0.5 \le h_0 \le 1$.
Hence, $r_*$ can reasonably be taken to be of ${\cal O} ({\rm pc})$.
Therefore
the study of static spherically symmetric solutions is important
and physically relevent even when a cosmological constant
$\Lambda$ is present.

 From an analysis of such static spherically symmetric solutions, we
find \cite{kcc} that for
BD theory, they are likely to be regular outside the
Schwarzschild horizon with no curvature singularities. However, for low
energy string theory, the presence of a non zero
cosmological constant leads to a curvature singularity, which is much
worse than a naked one as explained in the text. This singularity is
argued to persist when generic perturbations and higher order
string effects are included. However such naked singularities
have not been observed in our universe. Hence, one should require that
they be absent, atleast upto a distance $r_*$, upto which the
static spherically symmetric solutions analysed here are
expected to describe the gravitational field of point stars.
This will then impose a bound
$| \Lambda | < 10^{- 102} (\frac{r_*}{{\rm pc}})^{- 2}$
in natural units. Thus if $r_* \simeq 1 {\rm Mpc}$ then
$| \Lambda | < 10^{- 114}$, and if $r_*$ extends all the way upto the edge
of the universe ($10^{28} {\rm cm}$) then
$| \Lambda | < 10^{- 122}$ in natural units.

This paper is organised as follows. In section 2, the action and the
equations of motion for the graviton and the dilaton are given, for the
static spherically symmetric case. In section 3,
we consider the solutions
when $\Lambda = 0$, and analyse the PPN parameters and the singularities.
In section 4, $\Lambda$ is taken to be non zero. We show that for low
energy string theory, non zero $\Lambda$ leads to a naked curvature
singularity, and give arguments for its persistence when generic
perturbations and higher order string effects are included. In section 5,
we conclude with a summary.

\vspace{2ex}

\centerline{\bf 2. Equations of motion for graviton and dilaton}

\vspace{2ex}

Consider the following action for
graviton $(\tilde{g}_{\mu \nu})$ and  dilaton $(\phi)$ fields,
\begin{equation}\label{starget}
S = - \frac{1}{16 \pi \kappa}
\int d^4 x \sqrt{\tilde{g}} \, e^{\phi} \,
( \tilde{R} - \tilde{a} (\tilde{\nabla} \phi)^2 + \Lambda (\phi) )
\end{equation}
in the target space with coordinates
$ x^{\mu}, \; \mu = 0, 1, 2, 3$, where
$\kappa \; ( = 1$ in the following) is Newton's
constant. In our notation,
$R_{\mu \nu \lambda \tau} = \frac{\partial^2 g_{\mu \lambda}}
{\partial x^{\nu} \partial x^{\tau}} + \cdots$.
When $\tilde{a} = 1$, the action $S$ in equation (\ref{starget})
corresponds to the target space
effective action for low energy string theory, whose equations of motion
give the $\beta$-function equations for
$\tilde{g}_{\mu \nu}$ and $\phi$ in the sigma model approach to the
string theory.
$\Lambda (\phi)$ is the dilaton potential which, if constant, would
act as a tree level cosmological constant
in low energy string theory \cite{tdgang}, given by
$\Lambda = \frac{1}{2} (d + d_{int} - 10)$, which is zero
for a critical string and non zero for a non critical string.
The field $e^{- \frac{\phi}{2}}$ acts as a string coupling.
When $\tilde{a} = - \omega $ the above action corresponds to
Brans-Dicke (BD) theory, where $\omega > 0$ is the BD parameter.

In the effective action (\ref{starget}), which is written in
a frame (called physical frame in the following)
with metric $\tilde{g}_{\mu \nu}$, the curvature term is
not in the standard Einstein form. However, the standard form,
where the equations of motion are often easier to analyse,
can be obtained by a dilaton dependent conformal transformation
\[
\tilde{g}_{\mu \nu} = e^{- \phi} g_{\mu \nu}
\]
to the Einstein frame with metric $g_{\mu \nu}$.
The curvature scalars in these two frames are related by
\begin{equation}\label{rstring}
\tilde{R} = e^{\phi}
( R - 3 \nabla^2 \phi + \frac{3}{2} (\nabla \phi)^2 )
\end{equation}
where $\tilde{\,}$ refers to the physical frame.
The effective action now becomes
\begin{equation}\label{etarget}
S = - \frac{1}{16 \pi} \int d^4 x \sqrt{g} \,
( R + \frac{a}{2} (\nabla \phi)^2 + e^{- \phi} \Lambda (\phi) )  \; ,
\end{equation}
where $a \equiv 3 - 2 \tilde{a} \;  = 1$ for string theory and
$ = 2 \omega + 3$ for BD theory. The equations of motion for
$g_{\mu \nu}$ and $\phi$ that follow from this action, with
$\Lambda_{\phi} \equiv \frac{\partial \Lambda}{\partial \phi}$, are
\begin{eqnarray}\label{beta}
2 R_{\mu \nu} + a \nabla_{\mu} \phi \nabla_{\nu} \phi
+ g_{\mu \nu} \Lambda e^{- \phi}  & = & 0 \nonumber \\
a \nabla^2 \phi + (\Lambda - \Lambda_{\phi}) e^{- \phi} & = & 0 \; .
\end{eqnarray}

There is no specified form for the function
$\Lambda (\phi)$, either in Brans-Dicke theory or in string theory.
However, in the low energy limit of the string theory, the dilaton is
expected to acquire a mass, and consequently develop a potential of the
form $e^{- \phi} \Lambda (\phi) = m (\phi - \phi_0)^2 + \cdots$ around
the minimum of the potential. In two excellent papers \cite{hh},
the implications of such a massive dilaton have been thoroughly analysed
for static spherically symmetric solutions.
Hence, in the following we analyse only the case where
$\Lambda (\phi)$ is a constant, which corresponds to a tree level
cosmological constant in low energy string theory. Furthermore, since
$a \ge 1$ in string and BD theory, we also consider only $a \ge 1$.

We will look for static, spherically symmetric solutions
to equations (\ref{beta}). In the Schwarzschild gauge where
$d s^2 = - f d t^2 + f^{- 1} d \rho^2 + r^2 d \Omega^2, \;
d \Omega^2$ being the line element on an unit sphere, and where the fields
$f, \; r$, and $\phi$ depend only on $\rho$,
the equations (\ref{beta}) become
\begin{eqnarray}\label{rf}
\frac{(f r^2)''}{2} - 1 & = & ( f' r^2 )' \nonumber \\
= a ( \phi' f r^2 )' - \Lambda_{\phi} r^2 e^{- \phi}
& = & - \Lambda r^2 e^{- \phi} \nonumber \\
4 r'' + a r \phi'^2 & = & 0
\end{eqnarray}
where $'$ denotes $\rho$-derivatives.

Sometimes, it is more convenient
to work in the standard gauge where the line element is given by
$d s^2 = - f d t^2 + \frac{G}{f} d r^2 + r^2 d \Omega^2$,
and where the fields $f, \; G$, and $\phi$ depend only on $r$.
Equations (\ref{beta}) then become
\begin{eqnarray}\label{gf}
\frac{(f r^2)''}{2} - \frac{(f r^2)' G'}{4 G} - G
& = & ( f' r^2 )' - \frac{G' f' r^2}{2 G}
\nonumber \\
= a ( \phi' f r^2 )' - \frac{a \phi' G' f r^2}{2 G}
- \Lambda_{\phi} G r^2 e^{- \phi}
& = & - \Lambda G r^2 e^{- \phi} \nonumber \\
2 G' - a r G \phi'^2 & = & 0
\end{eqnarray}
where $'$ denotes $r$-derivatives now in the standard gauge.
The curvature scalar $\tilde{R}$ in the physical frame is given by
\begin{equation}\label{r}
\tilde{R} = \frac{(3 - a) f \phi'^2 e^{\phi}}{2 G}
+ \frac{(3 - 2 a) \Lambda}{a} - \frac{3 \Lambda_{\phi}}{a} \; .
\end{equation}
Using (\ref{gf}), it is easy to obtain the following equation for
$R_1 \equiv \frac{f \phi'^2 e^{\phi}}{G}$ :
\begin{equation}\label{r1}
R'_1 + (\frac{4}{r} + \frac{f'}{f} - \phi') R_1
= - 2 (\Lambda - \Lambda_{\phi}) \phi' \; .
\end{equation}
If $\Lambda$ is a constant then $\Lambda_{\phi} = 0$ and
the second equality in (\ref{gf}) can be integrated to obtain
\begin{equation}\label{phif}
\frac{f'}{f} - a \phi' = \frac{r_0 \sqrt{G}}{f r^2}
\end{equation}
where $r_0$ is an integration constant proportional to the mass of the star.

The metric can also be written in isotropic gauge where
the line element is given by
$d s^2 = - f d t^2 + F (d h^2 + h^2 d \Omega^2)$ where $f$ and $F$ are
functions of $h$ only. In this gauge, the observable
parameters of the metric $\tilde{g}_{\mu \nu}$ in the physical frame
can be extracted as follows. The mass of the star $M$ and
the relevent PPN parameters
$\beta$ and $\gamma$
are obtained \cite{will} by expanding the metric components
$\tilde{f}$ and $\tilde{F}$ in the physical frame, as
$h \to \infty$. These observables are defined by
\begin{eqnarray*}
\tilde{f} & = & 1 - \frac{2 M}{h} + \frac{2 \beta M^2}{h^2} + \cdots \\
\tilde{F} & = & 1 + \frac{2 \gamma M}{h} + \cdots \; .
\end{eqnarray*}
For Einstein's theory $\beta = \gamma = 1$. The physical
significance of the PPN parameters $\beta$ and $\gamma$ is that
$\beta$ measures the non
linearity in the superposition law of gravity, while $\gamma$
measures the space time curvature. Experimentally,
$\beta$ and $\gamma$ are obtained
by measuring the precession of the perihelia of the planets' orbits
and the time delay of radar echoes near the sun
respectively; their measured values are given by
$\frac{1}{3} (2 + 2 \gamma - \beta) = 1.003 \pm .005$ and
$\gamma = 1 \pm .001$ \cite{will}.

 From now on, we will take
$\Lambda (\phi) \equiv \Lambda = constant$ and $a \ge 1$.

\vspace{2ex}

\centerline{\bf 3. Solutions when $\Lambda$ is zero}

\vspace{2ex}

Consider first the solutions when $\Lambda = 0$. One then has the
standard Schwarzschild solution
\[
\tilde{f} = 1 - \frac{\rho_0}{\rho} \; , \; \;
\tilde{r} = \rho \; , \; \; \phi = \phi_0 \; ,
\]
where $\rho_0$ and $\phi_0$ are constants, which
describes the gravitational field of a point star
of mass $M = \frac{\rho_0}{2}$. There is a horizon at $\rho = \rho_0$
where $\tilde{g}_{tt} = \tilde{f} = 0$. The curvature scalar $\tilde{R}$
in the physical frame is regular everywhere, except at $\rho = 0$.
This is the well known black hole singularity and is hidden behind the
horizon. In the isotropic gauge, the solution becomes
\[
\tilde{f} = \left( \frac{1 - \frac{\rho_0}{4 h}}
{1 + \frac{\rho_0}{4 h}} \right)^2 \; , \; \;
\tilde{F} = \left( 1 + \frac{\rho_0}{4 h} \right)^4 \; ,
\]
where $h$ and $\rho$ are related by
\begin{equation}\label{hrho}
\rho = h \left( 1 + \frac{\rho_0}{4 h} \right)^2 \; .
\end{equation}
The PPN parameters are given by $\beta = \gamma = 1$ and are trivial.

However, there are also more general solutions \cite{b}-\cite{tcd6}
where the dilaton field $\phi$ and the PPN parameters are non trivial.
They are given, in the Schwarzschild gauge in the Einstein frame, by
\cite{b,tcd6}
\begin{eqnarray}\label{bdsoln}
f & = &
\left( 1 - \frac{\rho_0}{\rho} \right)^{\frac{1 - k^2}{1 + k^2}}
\nonumber \\
r^2 & = &
\rho^2 \left( 1 - \frac{\rho_0}{\rho} \right)^{\frac{2 k^2}{1 + k^2}}
\nonumber \\
e^{\phi - \phi_0} & = &
\left( 1 - \frac{\rho_0}{\rho} \right)^{\frac{2 l}{1 + k^2}}
\end{eqnarray}
where $k$ is a parameter and $l \equiv \frac{k}{\sqrt{a}}$. In the
physical isotropic gauge, the metric components become
\begin{eqnarray}
\tilde{f} & = &
\left( 1 - \frac{\rho_0}{\rho} \right)^{\frac{1 - k^2 - 2 l}{1 + k^2}}
\nonumber \\
\tilde{F} & = & \frac{\rho^2}{h^2}
\left( 1 - \frac{\rho_0}{\rho} \right)^{\frac{2 (k^2 - l)}{1 + k^2}}
\end{eqnarray}
where $h$ and $\rho$ are related as in (\ref{hrho}). Expanding the
functions $\tilde{f}$ and $\tilde{F}$ in inverse powers of $h$, as
$h \to \infty$, one gets the mass $M$ and the PPN parameters
$\beta$ and $\gamma$ as
\begin{eqnarray}\label{mass}
2 M & = & \frac{1 - k^2 - 2 l}{1 + k^2} \rho_0  \nonumber \\
\beta & = & 1 \nonumber \\
\gamma & = & 1 + \frac{2 l \rho_0}{(1 + k^2) M}  \; .
\end{eqnarray}
The parameter $\beta$ is trivial while $\gamma$ is non trivial if
$l \rho_0 \ne 0$.

The curvature scalar $\tilde{R}$ in the physical frame is given by
\begin{equation}\label{rtilde1}
\tilde{R} = \frac{\tilde{a} M^2 (\gamma - 1)^2 e^{\phi_0}}{\rho^4} \;
\left(1 - \frac{\rho_0}{\rho}\right)^{- \frac{1 + 3 k^2 - 2 l}{1 + k^2}}
\; .
\end{equation}
In the above equations $\rho_0$ is positive,
so that one obtains the standard Schwarzschild solution when $k = 0$.
Also the physical mass $M$, given by (\ref{mass}), must
be positive which then implies that $1 - k^2 - 2 l > 0$. Hence,
the metric component $\tilde{g}_{tt}$ in the physical frame vanishes at
$\rho = \rho_0$. The above condition on $k$,
discussed below in more detail, also implies that
$1 + 3 k^2 - 2 l >0$. Hence, the curvature scalar $\tilde{R}$
in (\ref{rtilde1}) becomes singular there, unless
$\gamma = 1$, {\em i.e.}\ unless the PPN parameters are trivial.
This singularity is naked, as will be shown presently.

We will first discuss the constraints on $k$.
The positivity of the physical mass $M$ in (\ref{mass}) implies that
$1 - k^2 - 2 l > 0$, which restricts the parameter
$k$ to be in the range
\begin{equation}\label{k1}
- \frac{1}{\sqrt{a}} - \sqrt{1 + \frac{1}{a}} < k <
- \frac{1}{\sqrt{a}} + \sqrt{1 + \frac{1}{a}} \; .
\end{equation}
However, the above equation turns out to be
only a weak constraint on $k$. A stronger one follows
requiring the PPN parameter $\gamma$ to lie within the experimentally
observed range $\gamma = 1 \pm .002$. In fact, from equations
(\ref{mass}), requiring $\gamma = 1 + \epsilon$ gives
\[
k = - \frac{1}{\sqrt{a}} \left( 1 + \frac{1}{\epsilon} \right)
\pm \sqrt{1 + \frac{1}{a} \left( 1 + \frac{1}{\epsilon} \right)^2} \; .
\]
Taking into the account the constraint on $k$ given by equation
(\ref{k1}), which implies that one should take the $+$ sign for the
square root above, we get
\[
k = - \frac{1}{\sqrt{a}} \left( 1 + \frac{1}{\epsilon} \right)
+ \sqrt{1 + \frac{1}{a} \left( 1 + \frac{1}{\epsilon} \right)^2}
\simeq - \frac{\epsilon \sqrt{a}}{2 (1 + \epsilon)} \; .
\]
Hence, if $\gamma$ is required to be such that
$| \gamma - 1 | \le |\epsilon|$, then one
gets the following stronger constraint on $k$:
\begin{equation}\label{k2}
|k| < \frac{|\epsilon| \sqrt{a}}{2 (1 + \epsilon)} \; ,
\end{equation}
where $|\epsilon| < .002$.

Now we will discuss the nature of the singularity at $\rho = \rho_0$.

\vspace{2ex}

\noindent 1. As can be seen from equation (\ref{rtilde1}), the
curvature scalar is singular at $\rho = \rho_0$; hence, this singularity
is not a coordinate artifact and cannot be removed by any coordinate
transformation.

\vspace{2ex}

\noindent 2. The metric on the surface $\rho = \rho_0$ has the signature
$0+++$, and hence, this surface is null and the singularity is a null
one.

\vspace{2ex}

\noindent 3. Consider an outgoing radial null geodesic, which
describes an outgoing photon. Since $d \tilde{s}^2 = 0$
for such a geodesic, its equation is given by
\[
\frac{d t}{d \rho} =
\left( 1 - \frac{\rho_0}{\rho} \right)^{\frac{k^2 - 1}{k^2 + 1}} \; ,
\]
where $t$ is the external time. This gives
\begin{equation}\label{nullgeo}
t = \rho_* + const
\end{equation}
where $\rho_*$, the analog of the `tortoise coordinate', is defined by
\[
\rho_* = \int d \rho
\left( 1 - \frac{\rho_0}{\rho} \right)^{\frac{k^2 - 1}{k^2 + 1}} \; .
\]
For $k = 0$, $\rho_* = \rho + \rho_0 \ln (\rho - \rho_0)$ is the
standard tortoise coordinate for Schwarzschild geometry, and it tends to
$- \infty$ as $\rho \to \rho_0$. For $k \ne 0$, $\rho_*$
given above cannot be explicitly evaluated for arbitrary $k$. However,
it can be shown that $\rho_*$ does not diverge as $\rho \to \rho_0$.
Near $\rho_0$, let $y = \rho - \rho_0 \; \to 0$. Then,
if $k \ne 0$ and obeys the bound given by equation (\ref{k2}), then
\[
\rho_* = \rho_0 \int d y \; y^{\frac{k^2 - 1}{k^2 + 1}} + \cdots \; , \;
\; = \frac{\rho_0 (k^2 + 1)}{2 k^2} y^{\frac{2 k^2}{k^2 + 1}} + \cdots
\]
where $\cdots$ denote higher order terms in $y$. The right
hand side of the above equation is finite as $y \to 0$, and thus
$\rho_*$ does not diverge as $\rho \to \rho_0$.

The outgoing radial null geodesic equation (\ref{nullgeo}) then implies
that a radially outgoing photon starting from $\rho_i \; ( \ge \rho_0 )$
at external time $t_i$ will reach an outside observer at
$\rho_f \; (\rho_i < \rho_f < \infty)$ at a finite external time
$t_f$ given by
\[
t_f - t_i = \rho_*(\rho_f) - \rho_*(\rho_i) \; .
\]
Since, as shown above, $\rho_*(\rho)$ has no divergence even when
$\rho = \rho_0$, it follows that a photon can travel from
arbitrarily close to the singularity to an outside observer
within a finite external time interval. Hence, the singularity at
$\rho = \rho_0$ is naked.

\vspace{2ex}

\noindent 4 a. Similarly a material particle
can also travel from arbitrarily
close to the singularity to an outside observer in a finite external
time interval. This can be shown as follows. Let the line element
be given by
\[
d \tilde{s}^2 = - g_0 d t^2 + g_1 d \rho^2 + g_2 d \Omega^2 \; ,
\]
where, for our case,
\[
g_0 = f e^{- \phi} \; , \; \;
g_1 = \frac{e^{- \phi}}{f} \; , \; \;
g_2 =  r^2 e^{- \phi}
\]
with $f, \;  e^{\phi}$, and $r^2$ given by equation (\ref{bdsoln}).
The corresponding geodesic equation for a material particle travelling
radially outward, which can be derived in a standard way as in
\cite{wein2}, is given by
\[
\frac{d t}{d \rho} = \sqrt{\frac{g_1}{g_0 (1 + E g_0)}} \; , \; \;
\frac{d \rho}{d \tau} = const \sqrt{\frac{1 + E g_0}{g_0 g_1}}
\]
where $\tau$ is the proper time (or equivalently the proper distance),
$E$ is the energy of the particle which is negative in our notation,
and $1 + E g_0 > 0$. For our case, these equations give
\begin{eqnarray*}
t & = & \int d \rho
\left( 1 - \frac{\rho_0}{\rho} \right)^{\frac{k^2 - 1}{k^2 + 1}} \;
\left( 1 + E \left( 1 - \frac{\rho_0}{\rho}
\right)^{\frac{1 - k^2 - 2 l}{k^2 + 1}} \right)^{- \frac{1}{2}}
+ const  \\
\tau & = & (const) \int d \rho
\left( 1 - \frac{\rho_0}{\rho} \right)^{\frac{- 2 l}{k^2 + 1}} \;
\left( 1 + E \left( 1 - \frac{\rho_0}{\rho}
\right)^{\frac{1 - k^2 - 2 l}{k^2 + 1}} \right)^{- \frac{1}{2}}
+ const \; .
\end{eqnarray*}
Since $(1 + E g_0) > 0$, the
only potential divergences in the above integrals are when
$\rho \to \rho_0$. However, analysing these integrals near
$\rho \to \rho_0$ as before, it can be seen that they do not diverge
as $\rho \to \rho_0$. Hence, just as in the case of a photon above, it
follows that a material particle can travel from arbitrarily
close to the singularity to an outside observer
within a finite external time interval. This again implies that
the singularity at $\rho = \rho_0$ is naked.

\vspace{2ex}

\noindent 4 b. By a similar analysis, it follows that
the proper distance $\tau$ between $\rho_f$ and
$\rho_i \; (\to \rho_0)$ is also finite.  The integral for
$\tau$ given above does not diverge since $1 + k^2 - 2 l > 0$,
which follows from the constraint $1 - k^2 - 2 l > 0$ discussed
before equation (\ref{k1}).

\vspace{2ex}

\noindent 5. The curvature scalar diverges as $\rho \to \rho_0$. The
ensuing tidal forces will rip away any physical apparatus as it nears
$\rho_0$. However, the information about this event can be communicated
to the outside observer in a finite external time since, as shown above,
a photon or a material particle can travel from arbitrarily close to the
singularity to an outside observer within a finite external time
interval.

For these reasons, the singularity at $\rho = \rho_0$ is naked and
physically  unacceptable. For recent detailed discussions on
naked singularities and their various general aspects, such as
their definition, physical unacceptability, various scenario for
their formation in Einstein's theory, etc.\ , see \cite{psj}).

We would like to make one further remark. The situation described here is
different from those corresponding to other solutions
in string theory where singular null horizons
appear. This is because, if and when the singularities do appear
for a charge neutral point star in the later case, they are always
hidden behind a horizon. For a point star with extremal charge, singular
null horizons can appear, but this situation again differs from the
present one in which only point stars with no charge
are considered. The naked,
singular `horizon' occurs in our case mainly because of the requirement
that the PPN parameter $\gamma$ for a charge neutral point star be
non trivial, {\em i.e.}\ $\gamma \ne 1$. The motivation for this
requirement has been discussed in the introduction.

One can gain more insight into the solution (\ref{bdsoln}) by comparing
it to that of \cite{gm,ghs}. Consider, as in \cite{gm,ghs},
a $U(1)$ gauge field $A_{\mu}$, coupled to
(\ref{etarget}) through the action
\begin{equation}
S_{em} = - \frac{1}{16 \pi} \int d^4 x \sqrt{g} \,
e^{\frac{k \phi}{\sqrt{a}}} F_{\mu \nu} F^{\mu \nu}
\end{equation}
where
$F_{\mu \nu} \equiv \partial_{\mu} A_{\nu} - \partial_{\nu} A_{\mu}$.
The general solution for the above system with the graviton, dilaton,
and a gauge field is given in the Schwarzschild gauge in the
Einstein frame, by \cite{gm,ghs}
\begin{eqnarray}
f & = & \left( 1 - \frac{\rho_1}{\rho} \right)
\left( 1 - \frac{\rho_0}{\rho} \right)^{\frac{1 - k^2}{1 + k^2}}
\nonumber \\
r^2 & = & \rho^2
\left( 1 - \frac{\rho_0}{\rho} \right)^{\frac{2 k^2}{1 + k^2}}
\nonumber \\
e^{\phi - \phi_0} & = &
\left( 1 - \frac{\rho_0}{\rho} \right)^{\frac{2 l}{1 + k^2}} \nonumber \\
F_{t \rho} & = & \frac{Q}{\rho^2}
\end{eqnarray}
where $l = \frac{k}{\sqrt{a}}$
and the remaining components of $F_{\mu \nu}$ are zero.
In the physical isotropic gauge, the metric components become
\begin{eqnarray}
\tilde{f} & = & \left( 1 - \frac{\rho_1}{\rho} \right)
\left( 1 - \frac{\rho_0}{\rho} \right)^{\frac{1 - k^2 - 2 l}{1 + k^2}}
\nonumber \\
\tilde{F}  & = & \frac{\rho^2}{h^2}
\left( 1 - \frac{\rho_0}{\rho} \right)^{\frac{2 (k^2 - l)}{1 + k^2}} \; ,
\end{eqnarray}
and, $h$ and $\rho$  are now related by
\[
\rho - \rho_0 = h (1 + \frac{\rho_1 - \rho_0}{4 h})^2 \; .
\]
Expanding the functions $\tilde{f}$ and $\tilde{F}$ in inverse
powers of $h$ as $h \to \infty$, one gets the mass $M$, the charge $Q$,
and the PPN parameters $\beta$ and $\gamma$ as
\begin{eqnarray*}
2 M & = & \rho_1 + \frac{1 - k^2 - 2 l}{1 + k^2} \rho_0 \\
Q^2 & = & \frac{\rho_1 \rho_0}{1 + k^2} \\
\beta & = & 1 + \frac{(1 - l) Q^2}{2 M^2} \\
\gamma & = & 1 + \frac{2 l \rho_0}{(1 + k^2) M}  \; .
\end{eqnarray*}
The parameter $\beta$ is non trivial if the charge $Q \ne 0$ while
$\gamma$ is non trivial if $l \rho_0 \ne 0$.

The curvature scalar $\tilde{R}$ in the physical frame is given by
\begin{equation}\label{rtilde2}
\tilde{R} = \frac{\tilde{a} M^2 (\gamma - 1)^2 e^{\phi_0}}{\rho^4} \;
\left(1 - \frac{\rho_1}{\rho}\right) \;
\left(1 - \frac{\rho_0}{\rho}\right)^{- \frac{1 + 3 k^2 - 2 l}{1 + k^2}}
\; .
\end{equation}
The metric component $\tilde{g}_{tt}$ in the physical frame vanishes at
$\rho = \rho_1$ and $\rho = \rho_0$. The curvature scalar
$\tilde{R}$ is regular at $\rho = \rho_1$ but,
since $1 + 3 k^2 - 2 l > 0$ for $a \ge 1$, it is singular
at $\rho = \rho_0$ unless $\gamma = 1$. This singularity is
hidden behind the horizon at $\rho_1$
if $\rho_1 > \rho_0$, and naked otherwise for the same reasons as given
following equation (\ref{rtilde1}).

Now, consider the charge neutral solution, {\em i.e.}\ $Q = 0$.
This can be obtained by setting
either $\rho_0 = 0$ or $\rho_1 = 0$. In the former
case, one gets the usual Schwarzschild solution with trivial values
for $\beta$ and $\gamma$. In the later case one gets the solution
described in (\ref{bdsoln}) where the parameter $\gamma$ is non trivial.

Thus, it can be seen from (\ref{rtilde1}) and (\ref{rtilde2}) that,
in BD or low energy string theory, a non trivial value for the parameter
$\gamma$ for a charge neutral point star implies the existence of a naked
singularity. Conversely, in these theories, the absence of naked
singularities necessarily implies that the PPN parameters $\beta$
and $\gamma$ for a charge neutral point star are trivial.
Thus if naked singularities are forbidden then, for such a
star, both the BD and the low energy string theory lead to the
same predictions for $\beta$ and $\gamma$ as in Einstein theory. In
particular, if the parameter $\gamma$ for such a star is found
to be different from one, then it cannot be explained by
either BD or low energy string theory, without implying the existence of
a naked singularity.  In that case an alternative
theory is needed that can predict a non trivial value for $\gamma$
for such a charge neutral point star, without any naked singularity.

\vspace{2ex}

\centerline{\bf 4. Solutions when $\Lambda$ is non zero }

\vspace{2ex}

Consider now the case when $\Lambda \ne 0$.
The equation involving $(\phi' f r^2)'$ in (\ref{gf})
is the equation of motion
for $\phi$ that follows from (\ref{etarget}).
However, this equation will be absent if the dilaton $\phi$ is absent.
Hence, in that case, this equation is to be ignored
and $\phi$ is to be set to zero in the remaining equations.
The solution to (\ref{gf}) is then given by
\[
f = 1 - \frac{r_0}{r} - \frac{\Lambda}{6} r^2 \; , \; \; G = 1 \; .
\]
The curvature scalar $\tilde{R} = \Lambda$. This solution
describes the static, spherically symmetric
gravitational field of a point star of mass $M = \frac{r_0}{2}$
in Einstein theory, in the presence of a cosmological constant
$\Lambda$ \cite{gh}.

In the presence of both the dilaton $\phi$, and the cosmological constant
$\Lambda$, the solution to equations (\ref{gf}) is not known in an
explicit form. Here we study this solution
and its implications.
The solution, required to reduce to the Schwarzschild one
when $\Lambda = 0$, would describe the static, spherically symmetric
gravitational field of a point star
in the graviton-dilaton system (\ref{starget}),
in the presence of a cosmological constant $\Lambda$.

For nonzero $\Lambda$, the following general features
are valid for any solution to equations (\ref{gf}):

(i)   The dilaton field $\phi$ cannot be a constant. In fact, the
only case where $\phi$ can be a constant for a non zero $\Lambda$
is when $\Lambda = \Lambda_{\phi}$, {\em i.e.}\
$\Lambda = \lambda e^{\phi}$. But, as can be seen from (\ref{etarget}),
this corresponds to pure Einstein theory with a cosmological constant
$\lambda$ and a free scalar field $\phi$.

(ii)  $\ln G$, and hence $G$, strictly increases since
$a \ge 1$ and consequently $(\ln G)' > 0$.

(iii) Consider the following polynomial
ansatz for the fields as $r \to \infty$.
\begin{eqnarray}
f & = & A r^k + \cdots \nonumber \\
G & = & B r^l + \cdots \nonumber \\
e^{- \phi} & = & e^{- \phi_0}  r^m + \cdots
\end{eqnarray}
where $\cdots$ denote subleading terms in the limit $r \to \infty$
(it can be easily shown that if one of the fields has an asymptotic
polynomial behaviour, then the others also have similar
behaviour). Substituting these expressions into equations (\ref{gf})
gives, to the leading order, $2 l = a m^2$ and
\begin{eqnarray}\label{asym}
\frac{(k + 2)}{2} (k + 1 - \frac{l}{2}) A r^k - B r^l
& = & k (k + 1 - \frac{l}{2}) A r^k  \nonumber \\
= - a m (k + 1 - \frac{l}{2}) A r^k
& = & - B \Lambda e^{- \phi_0} B r^{l + m + 2}  \; .
\end{eqnarray}
The last two equalities above imply $k = - a m  = l + m + 2$ which,
together with $2 l = a m^2$, lead to
$(m + 2) (m + \frac{2}{a}) = 0$. This gives the solution
$(k, l, m) = (2 a, 2 a, - 2)$ or $(2, \frac{2}{a}, - \frac{2}{a})$.
Using these relations and equations
(\ref{asym}) it follows that
\begin{eqnarray*}
k (k + 1 - \frac{l}{2}) A & = & - \Lambda e^{- \phi_0} B \nonumber \\
\left( (m + \frac{2}{a}) \Lambda e^{- \phi_0}
+ \frac{4}{r^{m + 2}} \right) B  & = & 0 \; .
\end{eqnarray*}

If $a > 1$, as in BD theory, then there is always a non trivial
asymptotic solution with non zero $A$ and $B$. For example,
$(k, l, m) = (2, \frac{2}{a}, - \frac{2}{a})$ and $B$ arbitrary.
Note that in the second relation above,
the term involving $r^{m + 2}$ can be ignored to the leading order,
since $m + 2 = 2 (1 - \frac{1}{a}) > 0$. Also, as can be easily checked
for this solution, the curvature scalar $\tilde{R}$ in the physical frame
is finite as $r \to \infty$.

However, if $a = 1$ as
in low energy string theory, then the above equations are consitent
only if $A = B = 0$. Hence, in this case, equations
(\ref{gf}) do not admit a non trivial solution where the fields are
polynomials in $r$ as $r \to \infty$.
A similar analysis will rule out the solutions where the fields have
polynomial-logarithmic behaviour asymptotically, {\em i.e.}\
where the fields behave as $r^m (\ln^n r) (\ln^p\ln r) \ldots$
to the leading order in $r$ as $r \to \infty$.

Thus, when $a > 1$, which includes BD theory, but not the low energy
string theory, a non trivial asymptotic solution for graviton and
dilaton exists asymptotically, as $r \to \infty$. Therefore it is very
plausible, although not proved here,
that a full solution can be constructed,
perhaps numerically, starting from a Schwarzschild solution near the
horizon and approaching the above asymptotic form as
$r \to  \infty$. The curvature scalar $\tilde{R}$ in the physical frame is
also likely to remain finite everywhere outside the Schwarzschild horizon.

However, for low energy string theory where $a = 1$, the situation is
totally different. To start with, no non trivial solution exists for
graviton and dilaton asymptotically as $r \to \infty$.
To further understand the solutions to (\ref{gf}),
we start with the Schwarzschild solution
and study how it gets modified when $\Lambda \ne 0$ (from now on
we set $a = 1$).
Then the expression involving $\Lambda$ in (\ref{gf}) acts as
a source for the fields $f, \; G$, and $\phi$, which can be solved
iteratively to any order in $\Lambda$. By construction, this would reduce
to the Schwarzschild solution in the limit $\Lambda \to 0$.
One thus gets
\begin{eqnarray}\label{fp}
f & = & 1 - \frac{r_0}{r} - \frac{\Lambda r^2}{6}
- \frac{\Lambda^2 r^4}{120} u_2
- \frac{4 \Lambda^3 r^6}{2835} u_3 + \cdots \nonumber \\
G & = & 1 + \frac{\Lambda^2 r^4}{72} v_2
+ \frac{2 \Lambda^3 r^6}{405} v_3 + \cdots \nonumber \\
\phi & = & \phi_0 - \frac{\Lambda r^2}{6} (1 + \frac{2 r_0}{r}
+ \frac{2 r_0^2}{r^2} \ln (r - r_0))   \nonumber \\
& & - \frac{\Lambda^2 r^4}{45} w_2
- \frac{197 \Lambda^3 r^6}{45360} w_3 + \cdots
\end{eqnarray}
where $\phi_0$ is a constant which can be set to zero without any physical
consequence, and $u_i, \; v_i, \; w_i$ are functions of $\frac{r_0}{r}$
and $\ln r$ which tend to $1$ in the limit $\frac{r_0}{r} \ll 1$.
Evaluating $u_i, \; v_i, \; w_i$ and/or further higher order terms will
not illuminate the general features of the solution. Also, the
series will typically have a finite radius of convergence beyond which
it is meaningless. Although it is possible to construct convergent
series in different intervals of $r$, it is difficult to extract
general features. Hence we follow a different approach.

It turns out that one can understand the general features
of the solutions using only (i) the equations (\ref{gf}),
(ii) the behaviour of the fields for small $r$, and (iii) their
non polynomial-logarithmic behaviour in the limit $r \to \infty$.

Note that $G = 1$ for Schwarzschild solution. Let $G$ has no pole
at any finite $r$. Then the requirement
that any solution to (\ref{gf}) reduce to the Schwarzschild one when
$\Lambda = 0$, combined with the fact that $G$ is a non decreasing
function, implies that $G (\infty)$, and hence, $B$ must be non zero.
Then the above analysis, which excludes polynomial behaviour for the
fields with non trivial coefficients, implies in particular, that
the fields cannot be constant, including zero, as $r \to \infty$.

Consider first the case where $r_0 = 0$. This will describe
the static, spherically symmetric gravitational field
of a star of negligible mass in low energy
string theory when $\Lambda \ne 0$. With $r_0 = 0$ and setting
$\phi_0 = 0$, equation (\ref{phif}) gives $e^{\phi} = |f|$.
It also follows from (\ref{fp}) that the function $f$
has a local maximum (minimum) at the origin if $\Lambda$ is positive
(negative). Away from the origin, the function $f$ can \\
(A) have no pole at any finite $r$ and go to either $\infty$
or a constant as $r \to \infty$, or \\
(B) have a pole at a finite $r = r_p$
(its behaviour for $r > r_p$ will not be necessary for our purposes).
We will also consider the case where \\
(C) $f$ has a zero at $r = r_H$.

Case A: The function $f$, and hence $G$, has no pole at finite $r$. From
the analysis preceding equation (\ref{fp}), it is already clear that
$f (\infty)$ cannot be a constant. This can also be seen as follows.
A necessary condition for $f (\infty)$ to be a constant is
that $f$ must have atleast one more critical point at
$0 < r_c \le \infty$. Let $f' (r_1) = 0$,
where $r_1 \le \infty$ is the first critical point after the origin
(note that $r_1 = \infty$ corresponds to the function $f$
decreasing (increasing) to a constant monotonically
if $\Lambda$ is positive (negative)).
Then it follows, from the behaviour of $f$ near the origin, that
$f$ must have a local minimum (maximum) at $r = r_1$, {\em i.e.}\
$f'' (r_1)$ must be positive (negative). This requirement holds good even
when $r_1 = \infty$. However, from equations (\ref{gf}) we get
\[
f'' (r_1) = - \Lambda e^{- \phi} G
\]
which is negative (positive) if $\Lambda$ is positive (negative).
This is in contradiction to the above
condition. Therefore $f' (r) \ne 0$ for any $r > 0$, including $r = \infty$.
Hence, the function $f$ obeying equations (\ref{gf}) and which
behaves as in (\ref{fp}) near the origin, cannot be constant
in the limit $r \to \infty$. From this, and the asymptotic non
polynomial-logarithmic behaviour of $f$, it follows that
$f (\infty) \to \infty$.

Whether these singularities are genuine or only coordinate
artifacts can be decided by evaluating the curvature scalar, $\tilde{R}$,
or equivalently $R_1 \equiv \frac{f \phi'^2 e^{\phi}}{G}$
which obeys the equation
\begin{equation}\label{r10}
R'_1 + \frac{4 R_1}{r} = - 2 \Lambda \frac{f'}{f}   \;   .
\end{equation}
See equations (\ref{r}), (\ref{r1}), and (\ref{phif}).
It can be seen that $R_1 (\infty)$ cannot be a constant.
For, if it were, then one gets
$f (\infty) \to r^{- \frac{2 R_1 (\infty)}{\Lambda}}$, a polynomial
behaviour for $f$ as $r \to \infty$, which is ruled out.
Equation (\ref{r10}) can be solved to give
\[
R_1 = - \frac{2 \Lambda}{r^4} \; \int dr \frac{r^4 f'}{f} \; .
\]
 From this it follows, as $r \to \infty$, that $\frac{f'}{f} > \frac{k}{r}$
for any constant $k$ (otherwise $R_1 (\infty) \to constant$).
This implies that $f$ grows faster than any power of $r$ when
$r \to \infty$. Evaluating the above integral in this limit, one
then gets $R_1 (\infty) \to \infty$.

Case B: The function $f$ has a pole at a finite $r = r_p < \infty$.
Then, from equation (\ref{r10}) it follows, near $r = r_p$,
that
\[
R_1 (r_p) = - 2 \Lambda \ln f (r_p) + {\cal O} (r - r_p)
\; \; \; \;  \to \; \; \pm \infty \; .
\]

Case C: The function $f$ has a zero at $r = r_H$. Then,
from equation (\ref{r10}) it follows, near $r = r_H$, that
\[
R_1 (r_H) = - 2 \Lambda \ln f (r_H) + {\cal O} (r - r_H)
\; \; \; \;  \to \; \; \pm \infty \; .
\]

Thus we see that $R_1$, and hence, the curvature scalar $\tilde{R}$
in the string frame, always diverges at one or more points
$r \equiv r_s = r_p, \; r_H, \; \infty$,
in low energy string theory when the cosmological
constant $\Lambda \ne 0$. These singularities, which will persist
even when $r_0 \ne 0$ as argued below, are naked.
In fact, they are much worse, as they are created by any object,
no matter how small its mass is. Thus at any
point of the string target space, there will be a singularity produced
by an object located at a distance $r_s$ from that point.

The above analysis also goes through when $r_0 \ne 0$ (the well known black
hole singularity present now at $r = 0$, independent of $\Lambda$ and hidden
behind the Schwarzschild horizon, will not concern us here). The easiest way
to see it is as follows. Let the radius of convergence of the series
in (\ref{fp}) be $\gamma$, {\em i.e.}\ the series converges for
$r < r_{con} \equiv \sqrt{\frac{\gamma}{|\Lambda|}}$ (the
expansion parameter in the series is $\Lambda r^2$). Thus, for
$r_0 \ll r_{con}$, its effect on the fields will be negligible by the
time $r$ is near $r_{con}$, and even more so beyond $r_{con}$,
as can be seen from (\ref{fp}), where the functions
$u_i, \; v_i, \; w_i \to 1$ in the limit $\frac{r_0}{r} \ll 1$.
Hence such a non zero $r_0$ will not affect
the poles and zeroes of $f, G$, and $\phi$ (which lie beyond $r_{con}$),
and therefore, the curvature singularites found before will persist.

Or, one can repeat the above analysis. Now, one does not start at
$r = 0$, where there is the well known black hole singularity
if $r_0 \ne 0$, but at some point beyond the horizon, where
the cosmological constant term,
$\frac{\Lambda r^2}{6}$, in the expression for $f$ in (\ref{fp})
dominates the mass term, $\frac{r_0}{r}$; that is, near when
$r^3 > \frac{6 r_0}{|\Lambda|}$. This value of $r$ can be ensured to
fall within the radius of convergence $r_{con}$ by choosing,
for a given non zero $\Lambda$,  a sufficiently small $r_0$, {\em i.e.}\
$6 r_0 < \sqrt{\frac{r^3}{|\Lambda|}}$. Then, the analysis proceeds as
before. If $\Lambda$ is positive (negative), then
the function $f$ will be decreasing (increasing), as $r$ is
increasing beyond the value $\frac{6 r_0}{|\Lambda|}$, where
the cosmological constant term in $f$ has started dominating the
mass term. One can then consider the cases (A), (B), and (C) as before,
and arrive at the same conclusion.

Thus, it is very likely that these singularities will also
persist for any $r_0$, since the restriction on $r_0$
above is only due to the limitation of our analysis.
The negligible effect of $r_0$,
in the presence of a cosmological constant, is also physically reasonable
since the cosmological constant can be thought of as vacuum energy
density and, as $r$ increases, the vacuum energy will overwhelm
any non zero mass of a star, which is proportional to $r_0$.

Similarly, one can consider a point star with charge $Q$. The fields
then will be modified by the presence of terms involving $\frac{Q^2}{r^2}$,
which will become negligible when $Q \ll r$. Thus, again by an analysis
similar to the above, the singularities can be shown to
persist even when the star is charged. Physically, the curvature
singularities arise because of the run away feed back effect of the
cosmological constant $\Lambda$ on the fields, as can be seen from
(\ref{gf}). Therefore, the effect worsens as $r$ increases. But,
the effect of mass, charge, etc.\ of a point star decreases as $r$
increases, cannot compensate for the effects of $\Lambda$,
and hence cannot remove the singularities arising due to
a nonzero $\Lambda$. From this, it is also clear that any generic
perturbation such as aspherical mass/charge distribution, non zero angular
momentum, etc.\ will not remove the above singularities either,
since the effects of these perturbations decrease with increasing $r$.

Thus we see that the static spherically symmetric
gravitational field produced by a star in low energy string theory
has a naked curvature singularity when the  cosmological
constant $\Lambda \ne 0$. The singularity is in fact much worse than
a naked one, and is stable under generic perturbations such as the
ones discussed above.

Now, as discussed in the introduction, the static spherically symmetric
solution describes the gravitational field of a spherical star
atleast upto a distance $r_* \simeq {\cal O} ({\rm pc})$,
in our universe regardless of its non static nature.
Therefore, the singularities described here must be absent atleast
upto a distance $r_*$.
This will then translate into a constraint on the cosmological
constant $\Lambda$, in the sigma model approach to low energy
string theory. If we take, somewhat arbitrarily, that the curvature
becomes unacceptably strong when $| \Lambda | r^2 \simeq 1$, then requiring
the absence of singularity upto a distance $r_*$ would give
\[
| \Lambda | r_*^2 < 1 \; ,
\]
which gives the bound
\[
| \Lambda | < 10^{- 102} (\frac{r_*}{{\rm pc}})^{- 2}
\]
in natural units. Thus if $r_* \simeq 1 {\rm Mpc}$ then
$| \Lambda | < 10^{- 114}$, and if $r_*$ extends all the way upto the edge
of the universe ($10^{28} {\rm cm}$) then
$| \Lambda | < 10^{- 122}$ in natural units.

The existence of the naked singularity in low energy string theory
when the cosmological constant,
$\Lambda \ne 0$ also means the following. If
$\Lambda$ was zero during some era in the evolution of the universe,
then the mechanism (if exists)
that enforces cosmic censorship - no evolution
of singularities from a generic, regular, initial configuration - would
also enforce the vanishing of $\Lambda$ in the long run, when the
universe would be evolving sufficiently slowly for the static
solutions to be applicable. Otherwise, cosmic censorship would be
violated by the singularities presented above.

We now remark on the validity of the low energy effective action
in string theory. This
action is only perturbative and will be modified by higher order
corrections in the regions of strong curvature. Hence, when these
corrections are included, the singularities
seen here may not be present.
However, these corrections will kick in only when the curvature is strong,
and the low energy effective action, and thus our analysis, is likely to
remain valid until then. Therefore while the fields and the curvature
may never actually become infinite, even when $\Lambda \ne 0$,
in the full string action with higher order corrections,
the present analysis indicates that
they will become sufficiently strong as to be
physically unacceptable, thus justifing the above conclusions.


\vspace{2ex}

\centerline{\bf 5. Conclusion}

\vspace{2ex}

We have analysed the static, spherically symmetric solutions to the \\
graviton-dilaton system, with or without electromagnetic couplings
and the cosmological constant. These solutions describe the
gravitational field of a point star.
The main results of the present analysis can be summarised as follows.

1. For a charge neutral point star,
neither BD nor low energy string theory
predicts non trivial PPN parameters, $\beta$ and $\gamma$,
without introducing naked singularities. Thus, if the naked
singularities are forbidden, then
these theories lead to the same predictions as in Einstein
theory in the static spherically symmetric regime.
In particular, if the parameter $\gamma$ for a charge neutral star is
observed to be different from one, then it cannot be explained by
either BD or low energy string theory, without implying the existence of
a naked singularity.

2. Upon coupling the cosmological constant $\Lambda$ as in the
action (\ref{starget}), in a way analogous to the coupling of a tree
level cosmological constant in low energy string theory, we find the
following for the static spherically symmetric solutions.
For BD type theories, these solutions are likely to exist with no
naked curvature singularities.
However, for low energy string theory, the
presence of a non zero cosmological constant leads to a curvature
singularity in the universe, which is much worse than a naked
singularity and is stable under generic perturbations.
As discussed before, the static spherically symmetric
solutions describe the gravitational field of a point star
atleast upto a distance $r_* \simeq {\cal O} ({\rm pc})$,
in our universe regardless of its non static nature.
Therefore, the singularities described here must be absent atleast
upto a distance $r_*$.  This implies a bound
$| \Lambda | < 10^{- 102} (\frac{r_*}{{\rm pc}})^{- 2}$ in natural units.
If $r_* \simeq 1 {\rm Mpc}$ then
$| \Lambda | < 10^{- 114}$, and if $r_*$ extends all the way upto the edge
of the universe ($10^{28} {\rm cm}$) then
$| \Lambda | < 10^{- 122}$ in natural units. We have also
argued that this result, and the consequent bound on $\Lambda$, are
unlikely to change even when the higher order string effects are included.

\vspace{2ex}

{\bf Note Added:} After the completion of our work, we
were informed by C. P. Burgess of reference \cite{burgess},
where spherically symmetric, six parameter family of four dimensional
string solutions have been studied.

\vspace{2ex}

Part of this work was carried out in School of Mathematics,
Trinity College, Dublin and was supported by Forbairt
SC/94/218. It is a pleasure to thank S. Sen for encouragement,
H. S. Mani and T. R. Seshadri for many discussions. We particularly
like to thank P. S. Joshi for numerous helpful communications regarding
various aspects of singularity, and the referee for
suggestions which, we believe, improved the quality and clarity
of our paper.



\begin{thebibliography}{999}
\bibitem{will}
C. M. Will, Theory and Experiment in Gravitational Physics,
Revised Edition, Cambridge University Press, 1993.
\bibitem{b}
C. H. Brans, Phys. Rev. {\bf 15} (1962) 2194.
\bibitem{gm}
G. W. Gibbons and K. Maeda, Nucl. Phys. {\bf B298} (1988) 741.
\bibitem{ghs}
D. Garfinkle, G. T. Horowitz, and A. Strominger,
Phys. Rev. {\bf D43} (1991) 3140.
\bibitem{ko}
R. Kallosh and T. Ortin, Phys. Rev. {\bf D48} (1993) 742;
T. Ortin, {\it ibid}, {\bf D47} (1993) 3136.
\bibitem{tcd6}
S. Kalyana Rama, Trinity College, Dublin preprint TCD-6-93.
\bibitem{gh}
G. W. Gibbons and S. W. Hawking, Phys. Rev. {\bf D15} (1977) 2738; \\
L. J. Romans, Nucl. Phys. {\bf B383} (1992) 395.
\bibitem{tdgang}
R. C. Myers, Phys. Lett. {\bf B199}, (1987) 371; \\
I. Antoniadis et al, Nucl. Phys. {\bf B328} (1989) 117; \\
M. Mueller, Nucl. Phys. {\bf B337} (1990) 37; \\
A. A. Tseytlin and C. Vafa, Nucl. Phys. {\bf B372} (1992) 443; \\
C. R. Nappi and E. Witten, Phys. Lett. {\bf B293} (1992) 309; \\
D. S. Goldwirth and M. J. Perry, preprint hepth/9308023, unpublished; \\
J. H. Horne and G. Horowitz, preprint NSF-ITP-93-95, YCTP-P17-93,
unpublished.
\bibitem{hh}
R. Gregory and J. A. Harvey, Phys. Rev. {\bf D47} (1993) 2411;
J. H. Horne and G. Horowitz, Nucl. Phys. {\bf B399} (1993) 169.
\bibitem{wilt}
S. Mignemi and D. L. Wiltshire, Class. Quant. Grav. {\bf 6} (1989) 987,
Phys. Rev. {\bf D46} (1992) 1475.  \\
D. L. Wiltshire, Phys. Rev. {\bf D44} (1991) 1100.
\bibitem{wein}
S. Weinberg, Rev. Mod. Phys. {\bf 61} (1989) 1.
\bibitem{kcc}
S. Kalyana Rama, Trinity College, Dublin preprint TCD-1-94,
to appear in Modern Physics Letters A.
\bibitem{wein2}
See, for example, S. Weinberg, {\em Gravitation and Cosmology:
Principles and Applications the General Theory of Relativity},
John Wiley, New York 1972; C. W. Misner, K. S. Thorne, and
J. A. Wheeler, {\em Gravitation}, W. H. Freeman and Company,
New York 1973.
\bibitem{burgess}
C. P. Burgess, R. C. Myers, and F. Quevedo, Preprint number
McGill-94/47, NEIP-94-011, hepth/9410142.
\bibitem{psj}
P. S. Joshi, {\em Global Aspects of Gravitation and Cosmology},
Oxford University Press, 1993.

\end{thebibliography}
\end{document}